\documentclass[superscriptaddress, reprint,amsmath,amssymb,showkeys]{revtex4-2}
\setcitestyle{super}
\usepackage{graphicx}
\usepackage{dcolumn}
\usepackage{bm}
\usepackage{color}
\usepackage{float}
\usepackage[utf8]{inputenc}

\usepackage{comment}
\usepackage{amsmath}
\usepackage{xr}
\usepackage{xr-hyper}
\usepackage{soul,xcolor}
\usepackage{ragged2e}
\usepackage{setspace}

\usepackage[colorlinks=true, urlcolor=blue, citecolor=blue, linkcolor=blue]{hyperref}

\soulregister\cite7
\soulregister\ref7
\setstcolor{blue}
\newcommand{\cmi}[1]{cm$^{-1}$}

\begin{document}

\title{Resolving Transient Electron-Phonon Coupling with Time-Resolved Spontaneous Raman Spectroscopy
}

\author{Guy Reuveni}
\author{Maya Levy Greenberg}
\affiliation{Department of Chemical and Biological Physics, Weizmann Institute of Science, Rehovot 7610001, Israel;\looseness=-1}
\author{Matan Menahem}
\affiliation{Department of Chemical and Biological Physics, Weizmann Institute of Science, Rehovot 7610001, Israel;\looseness=-1}
\author{Olle Hellman}
\email{olle.hellman@weizmann.ac.il}
\affiliation{Department of Molecular Chemistry and Material Science, Weizmann Institute of Science, Rehovot 7610001, Israel;\looseness=-1} 
\author{Omer Yaffe}
 \email{omer.yaffe@weizmann.ac.il}
\affiliation{Department of Chemical and Biological Physics, Weizmann Institute of Science, Rehovot 7610001, Israel;\looseness=-1}

\date{March 10, 2026}

\begin{abstract}
Understanding the interaction of charge carriers with lattice vibrations in the quasi-equilibrium regime is crucial for semiconductor functionality. However, the structural signatures of these interactions are often too subtle for conventional ultrafast techniques to detect. We developed a time-resolved spontaneous Raman technique based on time-correlated single-photon counting to track the spectral response following photoexcitation, providing sub-wavenumber spectral resolution and a few-hundred-picosecond temporal resolution. Unlike traditional pump-probe schemes, our method utilizes a modulated continuous-wave probe to maintain high spectral resolution, enabling detection of low-frequency Raman shifts down to $10~\mathrm{cm^{-1}}$. Applied to lightly boron-doped silicon, we resolve \textit{intra}-valence band and \textit{inter}-valence band electronic transitions. A coupled-mode analysis of transient phonon asymmetry, resulting from interference with the inter-valence band transitions, reveals electron-phonon coupling parameters that directly relate to carrier recombination. By capturing these subtle dynamical shifts,  we demonstrate that this platform offers a powerful probe for investigating electron--phonon interactions in long-lived excited states.

\end{abstract}

\maketitle


\section*{Introduction}
Following photoexcitation, hot carriers rapidly lose phase coherence and relax toward the band edges through carrier--carrier scattering and phonon emission on sub-picosecond timescales. After this initial relaxation, electrons and holes accumulate near the bottom of the conduction band and the top of the valence band, forming a quasi-equilibrium carrier distribution. The carrier population then gradually decreases through recombination over timescales ranging from picoseconds to microseconds~\cite{ShahJagdeep1999IRoP,binder1995nonequilibrium}.
This quasi-equilibrium stage is central to material functionality. During this period, processes such as transport of free carriers or polarons, self-trapping, and exciton dissociation occur through interactions between electronic and structural degrees of freedom~\cite{landsberg2003recombination,ToyozawaYutaka2003Osoe,ToyozawaYutaka2003Pats,ToyozawaYutaka2003Psc}.

Common ultrafast spectroscopies predominantly track interactions between electronic and structural degrees of freedom by monitoring observables related to electronic properties. Transient absorption and time-resolved photoluminescence follow carrier densities and band-structure evolution~\cite{KnowlesKathrynE2018Taou,kirchartz2020photoluminescence}, while terahertz spectroscopy probes carrier mobility and scattering dynamics~\cite{MORIYASU2024130139}. Coherent Raman techniques track impulsively driven vibrational modes, but their sensitivity diminishes once phase coherence is lost after a few picoseconds, limiting them largely to the hot-carrier cooling regime rather than the longer-lived quasi-equilibrium state~\cite{ShahJagdeep1999CSoS}.

In contrast, time-resolved spontaneous Raman scattering directly probes vibrational and structural dynamics throughout this quasi-equilibrium regime~\cite{Versteeg2018}. The temporal evolution of Raman frequency, linewidth, and intensity encodes changes in force constants, phonon lifetimes, and mode populations~\cite{Kash1985, Zhu2019DynamicalExcitations}. Because the Raman cross section depends on the electronic polarizability, the technique is inherently sensitive to transient modifications of electron–phonon coupling and local bonding~\cite{mills1970theory}. It therefore provides direct access to how vibrational energy redistributes across the lattice and how electronic excitations reshape the structural landscape~\cite{finardi2025transient,PhysRevLett.132.036902}.

The common approach to time-resolved spontaneous Raman scattering is a pump–probe scheme. An ultrafast pump pulse drives the system out of equilibrium, and another, time-delayed ultrashort probe pulse records the spontaneous Raman response \cite{Versteeg2018,Kash1985,Zhu2019DynamicalExcitations,finardi2025transient,Han2019,Zhu2018,Pellatz2021,Kang2008,yugami1986time,Oberli1987,Yan2009,Kash1987,Chou2024Ultrafast,raciti2025unraveling,Katsumi2023}. Clearly, the use of ultrashort pulses limits the spectral resolution of the Raman spectra. In principle, pump–probe Raman experiments can achieve high spectral resolution by using long probe pulses with narrow bandwidth. In practice, this is uncommon and even when the probe pulse is long and spectrally narrow, amplified spontaneous emission and residual spectral wings near the laser line make efficient filtering difficult~\cite{Versteeg2018}. These issues become especially severe when probing low-frequency Raman shifts below $\sim100~\text{cm}^{-1}$, where strong suppression of stray light is required. 

Here we implement an alternative approach to time-resolved Raman scattering based on time-correlated single-photon counting (TCSPC), in which individual Raman photons are time-tagged relative to the pump excitation~\cite{Madonini2021,Finlayson2021,kogler2020time}. By combining a modulated continuous-wave probe with TCSPC detection and a pulsed optical pump, we obtain a few-hundred-picosecond temporal resolution while maintaining high spectral resolution.
We demonstrate this approach in silicon, where photoexcitation induces two Raman-active valence-band transitions. An \textit{intra-valence-band} (intra-VB) transitions appears at low shifts up to $200~\mathrm{cm^{-1}}$~~\cite{Jain1976ElectronicSilicon,chandrasekhar1977intraband,Chandrasekhar1980Intra-P-Si,abstreiter2005light}, while an \textit{inter-valence-band} (inter-VB) transitions overlaps with the optical phonon at $521~\mathrm{cm^{-1}}$, creating an asymmetric Fano lineshape~\cite{Cerdeira1972EffectGe,Cerdeira1973EffectModes,cerdeira1973interaction,Zhu2019DynamicalExcitations,jouanne1985bound}. High spectral resolution enables a coupled-mode analysis to extract electron--phonon coupling parameters, which track the photoexcited carrier density during recombination.

\section*{Results and Discussion}
\subsection*{TCSPC-Based Time-Resolved Raman Experimental Platform}
\begin{figure}[]
    \centering
    \includegraphics[width=\linewidth]{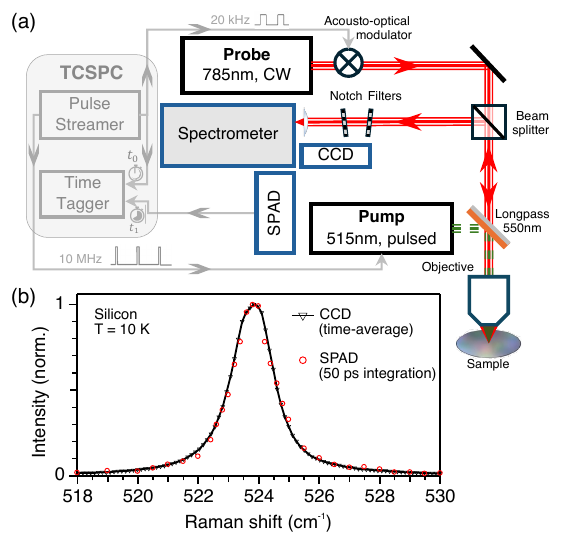}
    \caption{(a) Schematic of the time-resolved spontaneous Raman spectroscopy apparatus utilizing time-correlated single-photon counting. A modulated continuous-wave probe at 785~nm and a pulsed pump at 515~nm are co-aligned onto the sample. (b) Normalized probe-only Raman spectra of the silicon optical phonon at 10~K are recorded using a CCD detector (black) and reconstructed from a single 50~ps time bin using the single-photon avalanche photodiode with TCSPC (red). The overlap demonstrates that introducing picosecond time resolution does not compromise spectral resolution.}
    \label{fig:1_setup_10K}
\end{figure}

Time-resolved spontaneous Raman scattering measurements are performed using a setup based on TCSPC detection, schematically shown in Fig.~\ref{fig:1_setup_10K}(a). The instrument is built around a 1 m focal-length monochromator equipped with a 1800 grooves mm$^{-1}$ grating. The spectrometer is coupled to an optical microscope operating in a back-scattering configuration. A continuous-wave diode laser provides probe excitation at 785~nm. An acousto-optic modulator amplitude-modulates the probe beam at 20 kHz to generate an on–off sequence, enabling subtraction of pump-induced photoluminescence from the Raman signal. Non-equilibrium dynamics are initiated by a spatially co-aligned pulsed optical pump centered at 515 nm. The pump delivers 64 ps pulses with an energy of 2.3 nJ at a repetition rate of 10 MHz, corresponding to a fluence of approximately 70 mJ cm$^{-2}$.
Elastic scattering from the excitation line is suppressed using a beam splitter in combination with two volume holographic grating notch filters, enabling detection of low-frequency Raman shifts down to 10 cm$^{-1}$.

The spectrally dispersed Raman signal is collected at the monochromator's exit slit and detected by a single-photon avalanche photodiode (SPAD). Photon arrival times are recorded with a time-tagging module, with synchronization between the laser sources, modulation electronics, and detection system provided by a pulse generator. The overall temporal instrument response function, determined by the timing jitter of the spectrometer, detector, time tagger, and associated electronics, has a full width at half maximum of approximately 600~ps. A time-resolved Raman spectrum is obtained by sequentially scanning the monochromator grating. At each spectral position, TCSPC is used to accumulate a temporal histogram of Raman photon arrival times, from which the full time- and frequency-resolved Raman response is reconstructed.

The temporal instrument response function does not represent a fundamental limitation of the method. Improved temporal resolution is expected with state-of-the-art low-jitter avalanche photodiode arrays, which would also enable operation in spectrograph mode, thereby significantly reducing the acquisition time for a full spectrum~\cite{nissinen201816, tye2025time}. Additional improvements are anticipated with advanced time-tagging electronics and shorter pump pulse durations.
Figure~\ref{fig:1_setup_10K}(b) demonstrates that a spectrum reconstructed within a 50 ps temporal window retains the same spectral resolution as a time-averaged charge-coupled device (CCD) measurement. The silicon optical phonon is measured at 10 K under probe-only illumination, where its narrow linewidth provides a stringent test of spectral performance.

\begin{figure*}[]
    \centering
    \includegraphics[width=\textwidth]{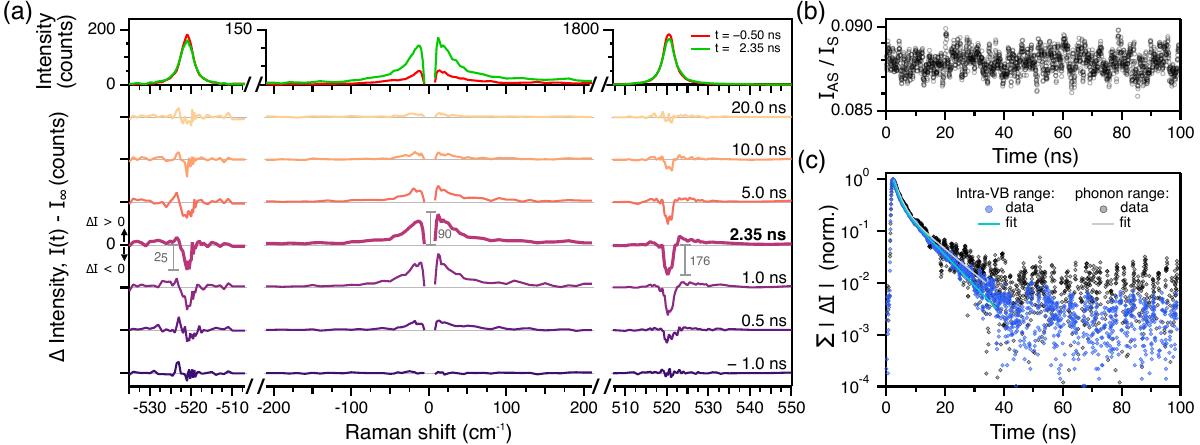}
   \caption{Time-resolved spontaneous Raman response of lightly boron-doped silicon at $280~\mathrm{K}$.
(a) Top panel: Raman spectrum of low-doped silicon pre-excitation (red) and post-excitation (green). Bottom panel: Differential spectra $\Delta I(\omega,t)$ obtained by subtracting the pre-excitation spectrum from each transient spectrum. The bold spectrum at $t = 2.35~\mathrm{ns}$ represents the maximum distortion. Differential spectra are offset for clarity, with gray axes indicating the zero level and gray scale bars representing the $\Delta I$ magnitude for each spectral range. The data reveal a transient amplification of the low-frequency \textit{intra-valence-band} (intra-VB) signal and a time-dependent modification of the optical phonon near $521~\mathrm{cm^{-1}}$ due to the \textit{inter-valence-band} (inter-VB) transitions.
(b) Time-dependent anti-Stokes/Stokes amplitude ratio. The constant ratio verifies that no significant lattice heating occurs during the measurement.
(c) Temporal evolution of the normalized integrated absolute differential intensity $\sum_{\omega} |\Delta I|$, integrated over the intra-VB ($10$--$200~\mathrm{cm^{-1}}$, blue) and phonon ($510$--$560~\mathrm{cm^{-1}}$, black) spectral ranges, shown with bi-exponential fits.}
    \label{fig:2_silicon}
\end{figure*}

\subsection*{Electronic and Phonon Dynamics in Photoexcited Silicon}

To demonstrate the capabilities of high-resolution time-resolved spontaneous Raman spectroscopy, we study the dynamics of photoexcitation of lightly boron-doped silicon wafer (dopant concentration $1\cdot10^{14}~\mathrm{cm^{-3}}$) with a native oxide layer as a model system. 

Figure~\ref{fig:2_silicon}(a) displays the Stokes and anti-Stokes transient Raman response for an estimated excess carrier density of approximately \(5 \cdot 10^{18}\) cm\(^{-3}\) during the first 0.5 ns after excitation at 280 K (refer to the Supplementary Information for similar data across a temperature range of 10 to 280 K). 
In the top panel, the spectrum before excitation (shown in red and averaged over the interval from -0.5 to 0 ns) is compared with the spectrum averaged over the period from 2.35 to 2.85 ns after excitation (shown in green). The bottom panel presents the differential spectra \(\Delta I(\omega,t) = I(\omega,t) - I_{\infty}\), where \(I_{\infty}\) refers to the spectrum prior to excitation. The most significant distortion occurs at \(t = 2.35\) ns, revealing two correlated features: a broad low-frequency Raman scattering signal that extends to approximately 200 cm\(^{-1}\), and a marked modification of the zone-center optical phonon at 521 cm\(^{-1}\).

The low-frequency response arises from intra-VB transitions of holes in the valence band of silicon, occurring at low Raman shifts up to approximately $200~\mathrm{cm^{-1}}$~~\cite{Jain1976ElectronicSilicon,chandrasekhar1977intraband,Chandrasekhar1980Intra-P-Si,abstreiter2005light}. Concurrently, a continuum of inter-VB hole transitions begins near $355~\mathrm{cm^{-1}}$ and extends to higher energies, overlapping with the optical phonon at $521~\mathrm{cm^{-1}}$~~\cite{Cerdeira1972EffectGe,Cerdeira1973EffectModes,cerdeira1973interaction,Zhu2019DynamicalExcitations,jouanne1985bound, Kanehisa1982InterbandP-silicon}. The interference between these transitions and the discrete phonon transition produces the characteristic asymmetric transient Fano lineshape.

Figure~\ref{fig:2_silicon}(b) demonstrates that these spectral changes are not caused by lattice heating. The anti-Stokes/Stokes amplitude ratio, which provides direct lattice thermometry~\cite{cooper2011raman,compaan1984resonance,Versteeg2018,Chou2024Ultrafast}, remains constant within experimental uncertainty, indicating that the lattice temperature does not increase during the transient response. Figure~\ref{fig:2_silicon}(c) shows the temporal evolution of the normalized, integrated absolute differential intensity, $\sum_{\omega} |\Delta I(\omega,t)|$, evaluated over the intra-VB Raman range ($10$--$200~\mathrm{cm^{-1}}$, blue) and the phonon range ($510$--$560~\mathrm{cm^{-1}}$, black). The latter is assigned to the optical phonon, the inter-VB transitions, and their mutual interference. Both signals rise promptly after photoexcitation and decay toward equilibrium, tracking the transient carrier population via a bi-exponential decay typical of carrier recombination in silicon~\cite{li2025applications,ghezellou2023role,dziewior1977auger}. Focusing on the dominant fast component, we extract decay constants of $\tau_1 = 1.8~\mathrm{ns}$ for the intra-VB signal and $\tau_1 = 2.7~\mathrm{ns}$ for the phonon feature. While these values differ, they are of the same order of magnitude and are consistent with effective carrier lifetimes in lightly doped silicon with a native oxide, where surface recombination dominates~\cite{MARKVART201327}.

\begin{figure*}[]
    \centering
    \includegraphics[width=\textwidth]{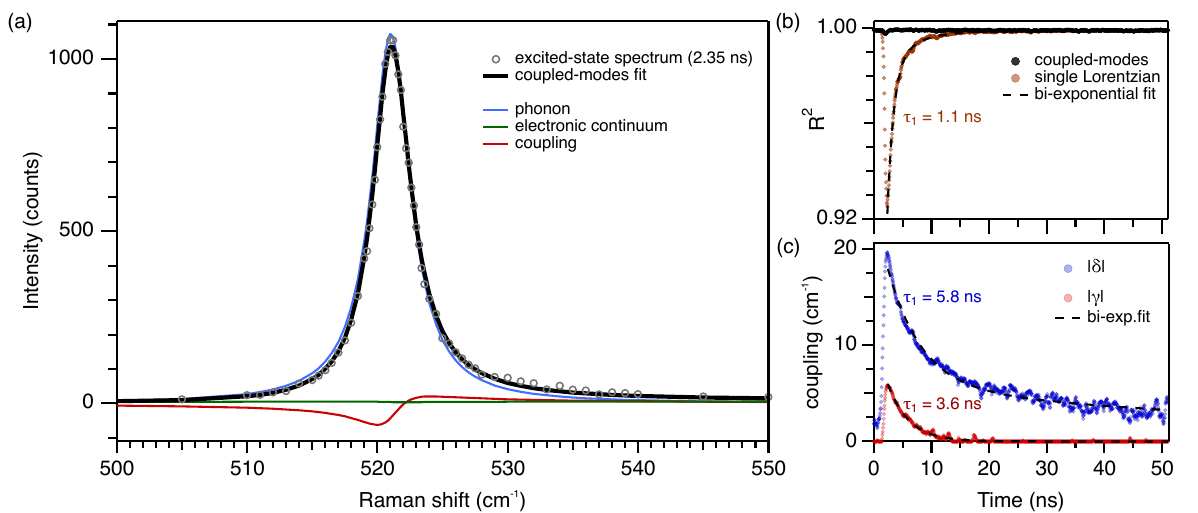}
   \caption{Time-resolved lineshape analysis of the optical phonon and the \textit{inter}-valence-band (inter-VB) spectral region in lightly boron-doped silicon at $280~\mathrm{K}$.
(a) Raman spectrum $2.35~\mathrm{ns}$ after photoexcitation. Gray circles indicate the time-averaged spectrum integrated over $2.35$--$2.85~\mathrm{ns}$, shown alongside the two-coupled-modes fit (black). The asymmetry of the peak is apparent compared to the purely phononic contribution (blue), with most of the asymmetry originating from the cross terms (red) between the inter-VB transitions and the phonon. The direct contribution from the inter-VB electronic transitions (green) is negligible.
(b) Statistical $R^{2}$ for the coupled-mode fit (black), compared with a single Lorentz oscillator fit (brown). 
(c) Temporal evolution of the absolute real ($|\delta|$, blue) and imaginary ($|\gamma|$, red) coupling parameters.
The dashed curves in (b) and (c) represent bi-exponential fits to the transients.
}
    \label{fig:TCM_fit}
\end{figure*}

\subsection*{Transient Electron–Phonon Interactions}
Next, we turn to the additional information encoded in the transient Raman lineshape, which becomes accessible owing to the high spectral resolution of the present measurements. We focus on the optical phonon and the associated inter-VB region, as the low-frequency intra-VB electronic response is spectrally broad and featureless, thus precluding a meaningful decomposition into distinct components.

Although the asymmetric phonon peak in the presence of photoexcited carriers is often described within the Fano formalism, we do not adopt that approach here. In the conventional Fano model, the discrete excitation is treated as a stationary eigenstate whose spectral width arises solely from its interaction with a continuum of electronic transitions~\cite{Fano1961EffectsShifts}. Within this framework, the phonon linewidth is rigidly tied to the strength of the coupling to the inter-VB transitions and does not account for the intrinsic phonon lifetime, which exists independently of the electronic continuum and exhibits a pronounced temperature dependence (see Supplementary Information). Moreover, the Fano formalism is constructed to diagonalize the energy matrix of stationary states and does not natively describe the time-dependent evolution of a coupled system. In our measurements, the phonon lineshape evolves continuously from an asymmetric profile shortly after excitation to a nearly symmetric form at longer delay times as the carrier population decays. This transient relaxation of the spectral asymmetry cannot be captured by a model based solely on static energy eigenvalues.

Instead, we employ a general description based on two coupled Lorentzian modes~\cite{scott1973soft}, consisting of a narrow mode representing the optical phonon and a broad mode representing the inter-VB transitions. This approach separates intrinsic phonon lifetime effects from coupling-induced distortions and provides a unified framework for analyzing the temporal dependence of the spectra. Formally, the system is described by a Green's function $G(\Omega)$~\cite{kwok1968green}, as previously implemented for coupled-mode systems~\cite{takagi1983coupled,benshalom2023phonon,Pinsk2025}. The Raman intensity is proportional to the corresponding spectral function, given by the imaginary part of the Green's function weighted by the Raman polarizability derivatives,
\begin{equation}\label{Greens_intensity}
I(\omega,T) \propto 
n(\omega,T)\sum_{ij}\left[-\mathrm{Im}\{G_{ij}(\omega,T)\}\chi_i \chi_j\right],
\end{equation}
where $\chi_1$ and $\chi_2$ denote the polarizability derivatives associated with the phonon and the inter-VB transitions, respectively. The interacting Green's function follows from the Dyson equation,
\begin{equation}\label{Dyson_eq}
G(\Omega)^{-1} = g^0(\Omega)^{-1} + \Sigma(\Omega),
\end{equation}
where $g^0(\Omega)$ is the harmonic Green's function and $\Sigma(\Omega)$ is the self-energy. For two coupled modes with frequency-independent parameters~\cite{takagi1983coupled,benshalom2023phonon}, these matrices take the form
\begin{equation}\label{hramonic_Green_func}
g^0(\Omega)^{-1}=
\begin{pmatrix}
\omega_1^2-\Omega^2 & 0 \\
0 & \omega_2^2-\Omega^2
\end{pmatrix},
\end{equation}
\begin{equation}\label{Sigma_self_energy}
\Sigma(\Omega,t)=
\Omega
\begin{pmatrix}
-i \Gamma_1 & \delta(t) + i \gamma(t) \\
\delta(t) + i \gamma(t) & -i\Gamma_2
\end{pmatrix}.
\end{equation}
Here $\Omega$ is the probing frequency, $\omega_1$ and $\omega_2$ are the bare frequencies of the phonon and inter-VB transitions mode, $\Gamma_1$ and $\Gamma_2$ are their respective damping constants, and $\delta(t)$ and $\gamma(t)$ represent the real and imaginary parts of the time-dependent intermode coupling.

To ensure physical robustness and avoid over-parameterization, we impose several constraints in the fitting procedure. The central frequency and width of the inter-VB transitions are fixed at $\omega_2 = 700~\mathrm{cm^{-1}}$ and $\Gamma_2 = 400~\mathrm{cm^{-1}}$, reflecting the broad inter-VB Raman response of silicon that overlaps the optical phonon near $521~\mathrm{cm^{-1}}$~~\cite{Cerdeira1973EffectModes,Kanehisa1982InterbandP-silicon}. The relatively modest phonon spectral distortion induced by photoexcitation supports keeping these parameters constant throughout the transient. The inter-VB transitions scattering strength $\chi_2$ is fixed at 20, as the fits are not sensitive to its precise value. The phonon frequency $\omega_1$, damping $\Gamma_1$, and scattering strength $\chi_1$ are fixed to the Lorentzian parameters extracted from the spectrum prior to the excitation. This procedure assumes that the transient modifications arise from coupling to the inter-VB transitions rather than from changes in the intrinsic phonon lifetime.

Under these constraints, the only free parameters are the coupling terms $\delta(t)$ and $\gamma(t)$. Physically, both are expected to scale with the photoexcited carrier concentration, since the effective electron--phonon interaction strength is proportional to the density of free carriers. We therefore treat them as time-dependent quantities that implicitly incorporate the carrier population. The real part $\delta$ couples the mode frequencies, while the imaginary part $\gamma$ couples their damping channels and represents a dissipative interaction~\cite{PhysRev.135.A1732}.

The results of this analysis are summarized in Fig.~\ref{fig:TCM_fit}. Panel (a) shows the coupled-modes fit to the excited-state spectrum, 2.35 nanoseconds after excitation, when the photoexcitation effect peaks. The cross term between the phonon and inter-VB transitions, shown in red, dominates the asymmetry, whereas the direct contribution of the inter-VB transitions (green) is negligible. Panel (b) presents the statistical $R^2$ of the coupled-mode fit, compared with that of a simple Lorentz oscillator model. The superior performance of the coupled-mode description highlights the transient contribution of the coupling to the inter-VB transitions and its temporal decay, demonstrating that the coupled-mode model is required to reproduce the asymmetric phonon profile. Panel (c) shows the temporal evolution of $\delta(t)$ and $\gamma(t)$. Dashed curves in (b) and (c) are the bi-exponential function fits to the transients.

The extracted decay constants ($\tau_1$) for $R^2(t)$, $\delta(t)$, and $\gamma(t)$ are not identical. This deviation from a single scaling factor reflects the fact that $\delta$ and $\gamma$ enter the Raman intensity expression in eq.~(\ref{Greens_intensity}) with different functional dependence, and the statistical measure $R^2$ is not expected to vary linearly with carrier concentration. Nevertheless, the overall agreement in their temporal behavior supports the interpretation that the transient lineshape evolution is governed by the decay of the photoexcited carrier population.

In summary, high-resolution time-resolved spontaneous Raman spectroscopy reveals the dynamics of electron--phonon interactions in photoexcited silicon during the quasi-equilibrium regime. A coupled-mode analysis shows that the transient asymmetry of the optical phonon arises from its time-dependent interaction with the inter-VB transitions. The coupling parameters $\delta(t)$ and $\gamma(t)$ quantify how the photoexcited carrier population modifies the phonon properties, and their decay directly follows the reduction of the carrier density during recombination.

\section*{Conclusions}
Our work addresses the challenge of resolving the subtle interactions between charge carriers and lattice vibrations during the quasi-equilibrium regime of semiconductors. By implementing a time-correlated single-photon counting (TCSPC) approach with a modulated continuous-wave probe, we achieve sub-wavenumber spectral resolution and a few-hundred-picosecond temporal resolution. This allows for the detection of low-frequency shifts and high-frequency phonon modifications.
We demonstrated the effectiveness of this technique in lightly boron-doped silicon, resolving the transient amplification of the \textit{intra}-valence-band transitions and the time-dependent interference between the optical phonon and the \textit{inter}-valence-band transitions. Using a coupled-mode analysis, we extracted time-dependent coupling parameters, $\delta(t)$ and $\gamma(t)$, which directly track the carrier recombination dynamics.
This experimental platform offers a versatile high-resolution probe for investigating the complex electron-phonon interactions that define the functional properties of modern semiconductors.

\subsection*{Author Contributions}
O.Y. and G.R. conceived the study. G.R. and M.L.G. developed the optical setup and performed the measurements. G.R. carried out the data analysis. O.H. developed the theoretical models. M.M. contributed to the setup development and numerical implementation of the fitting procedures. O.Y. supervised all stages of the project. All authors contributed to the interpretation of the results and to the final manuscript.

\subsection*{Acknowledgments}
The authors thank Yehiam Prior and Maor Asher for their insightful feedback and Lior Segev for software development.
G.R. and M.L.G. acknowledge support for this research by the Institute for Environmental Sustainability of the Weizmann Institute of Science.
O.Y. acknowledges funding from the European Research Council (850041 — ANHARMONIC).



%


\clearpage
\vspace{10em}

\renewcommand{\bibnumfmt}[1]{[S#1]}  
\renewcommand{\citenumfont}[1]{S#1}  

\renewcommand{\thepage}{S\arabic{page}}  
\renewcommand{\thesection}{S\arabic{section}}   
\renewcommand{\thesubsection}{S\arabic{section}.\alph{subsection}} 
\renewcommand{\thetable}{S\arabic{table}}   
\renewcommand{\thefigure}{S\arabic{figure}}
\renewcommand{\theequation}{S\arabic{equation}}
\setcounter{page}{1}
\setcounter{figure}{0}
\setcounter{equation}{0}

\clearpage
\onecolumngrid
\setstretch{2}

\begin{center}
    {\Huge \textbf{Supplementary Information}} \\[2.5em]
    {\Large \textbf{Resolving Transient Electron-Phonon Coupling with \\[0.25em] Time-Resolved Spontaneous Raman Spectroscopy}} \\[2.5em]
    
    Guy Reuveni,$^{1}$ Maya Levy Greenberg,$^{1}$ Matan Menahem,$^{1}$ Olle Hellman,$^{2,*}$ and Omer Yaffe$^{1,\dag}$\\[1em]
    
    $^1$\textit{Department of Chemical and Biological Physics,} \\
    \textit{Weizmann Institute of Science, Rehovot 7610001, Israel;} \\[0.5em]
    $^2$\textit{Department of Molecular Chemistry and Material Science,} \\
    \textit{Weizmann Institute of Science, Rehovot 7610001, Israel;} \\[1.5em]
    
    (Dated: March 10, 2026)
\end{center}

\vspace*{\fill} 
\noindent\rule{1.5in}{0.4pt} \\ 
{\footnotesize 
$^*$ \href{mailto:olle.hellman@weizmann.ac.il}{olle.hellman@weizmann.ac.il} \\[0.2em]
$^\dag$ \href{mailto:omer.yaffe@weizmann.ac.il}{omer.yaffe@weizmann.ac.il}
}

\onecolumngrid

\clearpage

\section{Experimental}
\subsection{TCSPC Time-Resolved Raman Experimental Platform}
The time-resolved Raman scattering measurements were carried out using a setup developed by extending our existing high-resolution Raman platform~\cite{asher2020anharmonic, menahem2021strongly} (Figure 1 in the main article). The system is built around a 1 m focal-length monochromator (FHR 1000, Horiba), with a 1800 grooves mm$^{-1}$ grating. 
The Raman probe laser is a 785 nm continuous-wave diode laser (XTRA II, Toptica Inc., USA), which is spectrally cleaned with amplified spontaneous emission filters and subsequently spatially filtered.
The probe beam is directed into an acousto-optical modulator (M1206-P110L-1, ISOMET), where the first order, amplitude-modulated diffracted beam is taken downstream towards the sample, producing an on/off sequence at 20 kHz.
The 515 nm pulsed pump laser (LDH-P-FA-515L, PicoQuant) with 2.3 nJ, 64 ps pulses at 10 MHz is passed through a telescope to ensure that the pump and probe focal points are precisely overlapping at the sample surface.
The pump beam is then merged with the central optical path via a dichroic mirror (DMLP550R, Thorlabs), spatially co-aligned with the probe beam, and focused onto the same spot on the sample. 
Both pump and probe lines are focused on the sample through an optical microscope with 0.55NA/50x objective (LD EC Epiplan-Neofluar 50x/0.55 DIC M27, Zeiss) in a back-scattered configuration, passing through a half-wave plate (Thorlabs) to align the excitation polarization with the crystal axis to maximize the equilibrium Raman intensity.
The sample is measured in a cryogenic-cooled cryostat (Janic Inc.) under high vacuum.
Elastic scattering from the excitation line is suppressed using two volume holographic grating notch filters with OD$>$4 (SureBlock, Coherent) in combination with a beam splitter (NoiseBlock, Coherent), enabling reliable detection of Raman shifts down to 10 cm$^{-1}$ from the laser line.

The spectrally dispersed Raman signal is collected at the width-variable slit of the monochromator and detected using either a CCD (Synapse+, Horiba) or a single-photon avalanche photodetector (Excelitas, SPCM-AQRH-46) with additional exit slit filtering the bandwidth that enters it. Photon arrival times to the SPAD are recorded with a time-tagging module (Ultra 4 Value, Swabian Instruments), with synchronization between the laser sources, modulation electronics, and detection system provided by a Pulse Streamer 8/2 (Swabian Instruments). The time tags are recorded in 50~ps time bins, and the overall instrument temporal response function has a full-width at half-maximum of 600~ps, derived from the timing jitter of the pump laser, the electronics, the SPAD, and the diffraction grating.

A complete time-resolved Raman spectrum is obtained by sequentially scanning the monochromator grating. At each spectral position, time-correlated single-photon counting is used to accumulate a temporal histogram of Raman photon arrival times for 20 minutes, from which the full time- and frequency-resolved Raman response is reconstructed.
Average laser power used for all measurements varied between 12-30 mW for the probe laser and 12-14 mW for the pump laser.

\subsection{Silicon sample}
The Silicon sample is a \(<100>\) cut, Boron-doped with resistivity 70-85 $\Omega$~cm which translates into dopant concentration of $\sim1\cdot10^{14}$ atoms cm$^{-3}$ (Virginia Semiconductors). The Silicon wafer has a native oxide layer.

\section{Calculation of     the Time-Dependent Carrier Density}
To model the spatial evolution of the charge-carrier distribution, we utilized a three-dimensional Green's function approach representing the diffusion of $N_0 = 1 \cdot 10^9$ initially excited holes, determined by the absorbed pump-pulse energy in silicon. This corresponds to an initial effective charge-carrier density of approximately $5\cdot10^{18}~\mathrm{cm^{-3}}$ within the first $\sim0.5~\mathrm{ns}$ after excitation. The spatial distribution $n(r, z, t)$ of these carriers diffusing from a localized point source in the silicon substrate is defined by:
\begin{equation}
    n(r, z, t) = \frac{N_0}{(4\pi Dt)^{3/2}} \exp\left( -\frac{r^2 + (z - z_0)^2}{4Dt} \right)
\end{equation}
where D = 12 cm$^2$~s$^{-1}$ is the diffusion coefficient~\cite{mohammad1993temperature}, $r$ and $z$ are the radial and axial coordinates, and $z_0$ represents the initial excitation center. This center is defined by half the absorption depth of the 515~nm pump beam ($z_0 = 0.5~\alpha_{515}^{~~~-1}$), where $\alpha_{515} = 9.25 \cdot 10^3~\text{cm}^{-1}$~~\cite{Green1995Optical}, yielding an initial excitation depth of $z_0 \approx 0.54~\mu\text{m}$.

The observable carrier density is calculated as an average over the effective probe volume $V_{\text{probe}}$, modeled using the probe-laser wavelength, microscope objective numerical aperture, and the refractive index of silicon. 
The instantaneous average density $\langle n(t) \rangle$ is thus determined by numerically integrating the Gaussian density distribution over a conical geometry representing the probe volume:
\begin{equation}
    \langle n(t) \rangle = \frac{1}{V_{\text{probe}}} \iiint_{V_{\text{probe}}} n(r, z, t) \, dV
\end{equation}

\section{data analysis}

\subsection{Preliminary data treatment - background photoluminescence and noise subtraction}

\begin{figure}[h!]
    \centering
    \includegraphics[width=\linewidth]{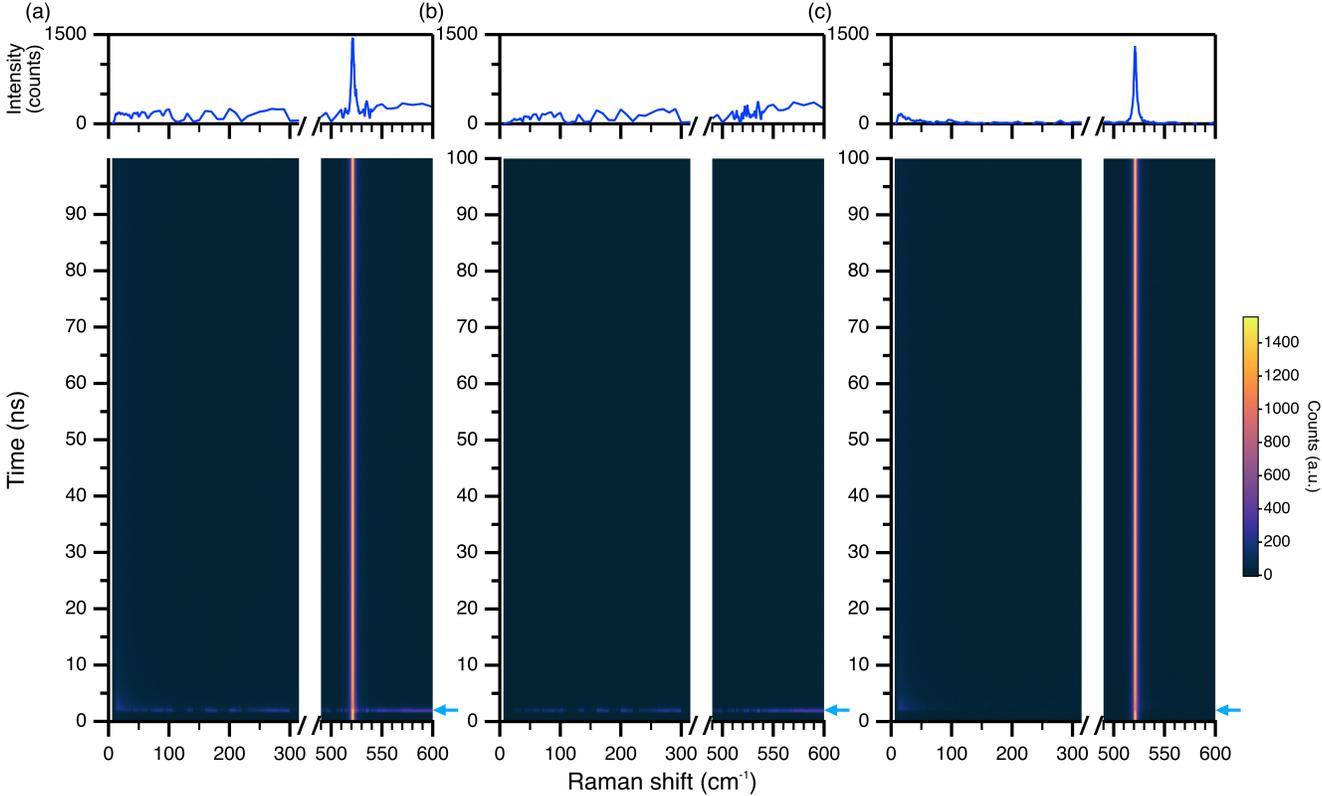}
    \caption{\normalsize \justifying Time-resolved Raman measurement scheme demonstrates the efficiency of the photoluminescence and background subtraction.
    Raw data at 280~K are presented as false-color heatmaps, with brighter colors indicating higher intensity. 
    The time-resolved Raman data is collected under (a) simultaneous illumination of the pump and probe lasers and (b) under pump laser only, when the probe laser amplitude was off-modulated at the AOM. (c) The subtraction of the pump-only matrix from the pump+probe matrix yields the time-resolved Raman matrix that is used as the starting point for all subsequent analyses in this paper.
    Top panels - single Raman spectra reconstructed at the 2.00~ns time bin, marked with blue arrows on the heatmaps.}
    \label{fig_SI:raw_data_and_subtracted_matrices}
\end{figure}

\begin{figure}[h!]
    \centering
    \includegraphics[width=\linewidth]{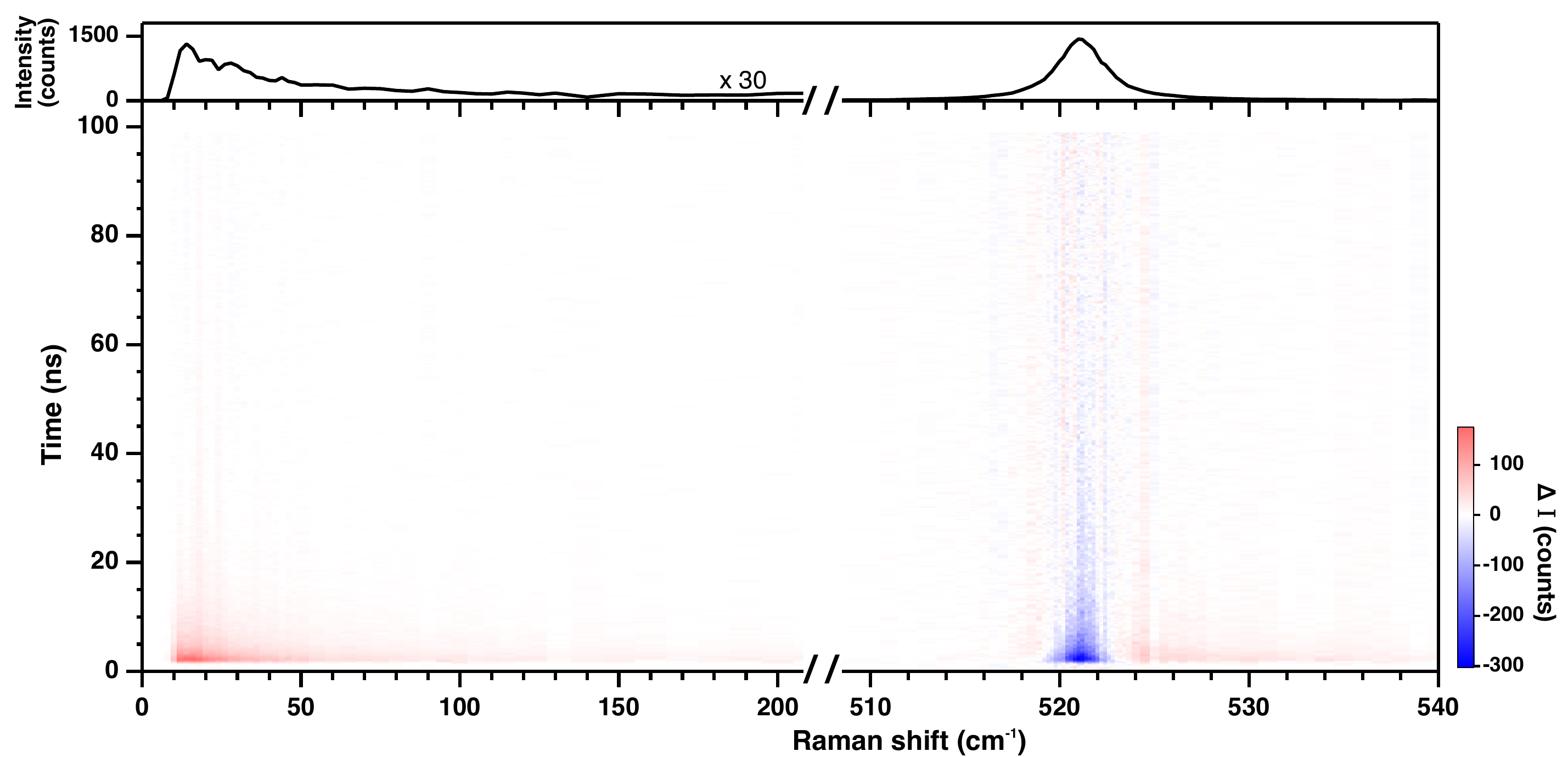}
    \caption{\normalsize \justifying Time-resolved spontaneous Raman response of lightly boron-doped silicon. Heatmap of the Stokes differential Raman intensity $\Delta I(\omega,t)$ at 280~K, defined relative to the pre-excitation spectrum (top), showing a transient low-frequency electronic continuum and a time-dependent modification of the optical phonon near 521~cm$^{-1}$.}
    \label{fig_SI:diff_heatmap_280K}
\end{figure}

The raw data consist of two time-resolved Raman-spectra matrices, corresponding to the on/off modulated probe signal: one acquired with both pump and probe beams (fig.~\ref{fig_SI:raw_data_and_subtracted_matrices}(a)), and one with the pump only (fig.~\ref{fig_SI:raw_data_and_subtracted_matrices}(b)). 
Subtracting the pump-only matrix from the pump+probe matrix removes photoluminescence and other pump-induced backgrounds, as shown in the spectra at the top of the two matrices, which represent a spectrum at a specific time bin of 2.00-2.05 ns.
The resulting difference matrix (fig.~\ref{fig_SI:raw_data_and_subtracted_matrices}(c)) isolates the Raman background present at equilibrium, along with any additional modifications of that background not directly attributable to the pump excitation, such as electron-phonon interaction effects on the phonon lineshape.
These modifications in the difference matrix are not visible to the naked eye, here specifically because they are very subtle.
To that end, we define a time-smoothing window of 10 spectra, each assigned to 50 ps, and, in total, equivalent to 500 ps, and apply a moving average throughout the time-dependent data.
The time-averaged difference matrix serves as the starting point for all subsequent analyses.
We extract the spectrum in the last time bin of the averaged difference matrix, referred to as the pre-excitation spectrum, and subtract it from all other time-dependent averaged spectra to get the differential spectra matrix (fig.~\ref {fig_SI:diff_heatmap_280K}).

\subsection{temperature-dependent integrated differential intensity }

To quantify and compare the dynamics of the photoexcited effect on the two electronic transitions and the effect on the optical phonon in the studied temperature range, each differential intensity spectrum $\Delta I(\omega,t)$ was integrated over the two spectral regions associated with the carrier-induced response, namely the low-frequency \textit{intra}-valence-band (intra-VB) range ($10$–$200~\mathrm{cm^{-1}}$) and the optical phonon range ($510$–$560~\mathrm{cm^{-1}}$) that is modified by interference with the \textit{inter}-valence-band (inter-VB) transitions.
The full time-resolved, background-subtracted differential Raman spectra matrix at 280~K is shown in fig.~\ref{fig_SI:diff_heatmap_280K} as a false-color heatmap, where red and blue indicate signal enhancement or suppression, respectively, compared to the pre-excitation spectrum, shown in the top panel. 
This figure shows an example of the data used to extract the transient absolute integrated intensity. It shows a prompt response right after excitation ($t=0$ ns), followed by an exponential decay that almost vanishes within 30~ns, returning to the pre-excitation spectrum.
The low-frequency electronic Raman signal is amplified by the elevated density of excited charge carriers, which scatter light inelastically.
The zone-center phonon around $521~cm^{-1}$ is distorted by coupling to the excited inter-VB transitions, as explained in the main article, showing an intensity quench at the peak (dark blue) and an increasing asymmetric signal on its higher-frequency side (light red).

\begin{figure}[h!]
    \centering
    \includegraphics[width=\linewidth]{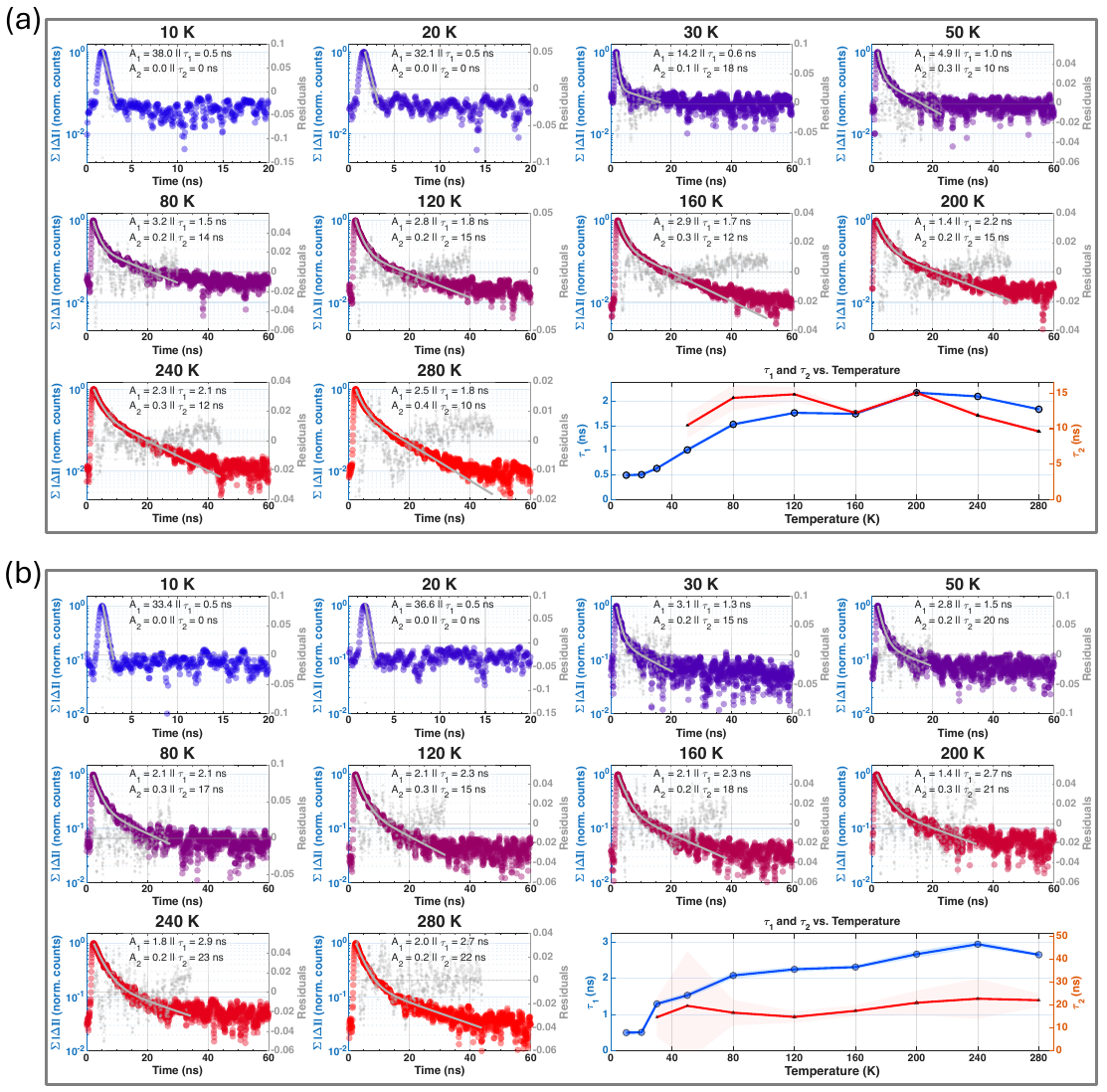}
    \caption{\normalsize \justifying Temperature-dependent integrated absolute differential intensity $\sum_{\omega} |\Delta I(\omega,t)|$ of (a) the intra-VB range and (b) the phonon range. Gray curves are the bi-exponential decay fit per transient, and gray circles are the fit residuals. The amplitude and decay constants are annotated next to the transients, and summarized at the bottom-right corners with the fast and slow decay constants plotted in blue and red, respectively. Shading represents the fit error of the decay constants, and the relative marker size represents the ratio of the fast and slow exponential amplitudes at each temperature.}
    \label{fig_SI:Tdep_int_diff_int}
\end{figure}

The integrated differential intensity transient at each temperature was background-subtracted to eliminate different noise levels between temperatures, normalized, and fitted to a bi-exponential decay function, as shown in fig.~\ref{fig_SI:Tdep_int_diff_int} in gray curves; its results are annotated next to each decaying curve.
The transients were fitted in the time range between the peak value and the first reaching the equilibrium value.
The resulting temperature-dependent decay constants $\tau_1$ and $\tau_2$ are presented in the bottom-right corner of each subfigure. 
At lower temperatures ($\leq$~30~K), the slow component is negligible, and only one decay constant is effectively dominant.

\newpage


\subsection{Lineshape analysis}
\subsubsection{Fano resonance model}
The Fano interference model describes a system in which a discrete scattering state, the optical phonon, interacts through energy exchange with a continuum background of electronic transitions, here, the inter-valence band hole transitions \cite{Fano1961EffectsShifts, Burke2010RamanSilicon}.
The temperature- and time-dependent data were fitted to a Fano lineshape of the form:
\begin{subequations}\label{Fano}
    \begin{equation}
        I = I_0 \frac{(q+\varepsilon)^2}{1+\varepsilon^2}
    \end{equation}
    \begin{equation}
        \varepsilon = \frac{\omega-\omega_0}{\frac{\Gamma}{2}}
    \end{equation}
\end{subequations}

where $I_0$ is an intensity prefactor, $q$ the Fano asymmetry parameter, $\omega$ the measured frequency, $\omega_0$ the resonance frequency, and $\Gamma$ the squared matrix element of the coupling between the continuum and the discrete states~\cite{Fano1961EffectsShifts, cerdeira1973interaction}, experimentally correlated with the peak width.
Large $|q|$ values yield a nearly symmetric lineshape, while smaller values produce pronounced asymmetry.
$q$ is proportional to the ratio of the Raman tensors for phonon and electronic scattering:
\begin{equation}\label{q^2Gamma} 
    \Gamma q^2 \propto \left|\frac{R_{phonon}}{R_{electronic}}\right|^2 
\end{equation}

\begin{figure}[h!]
    \centering
    \includegraphics[width=0.88\linewidth]{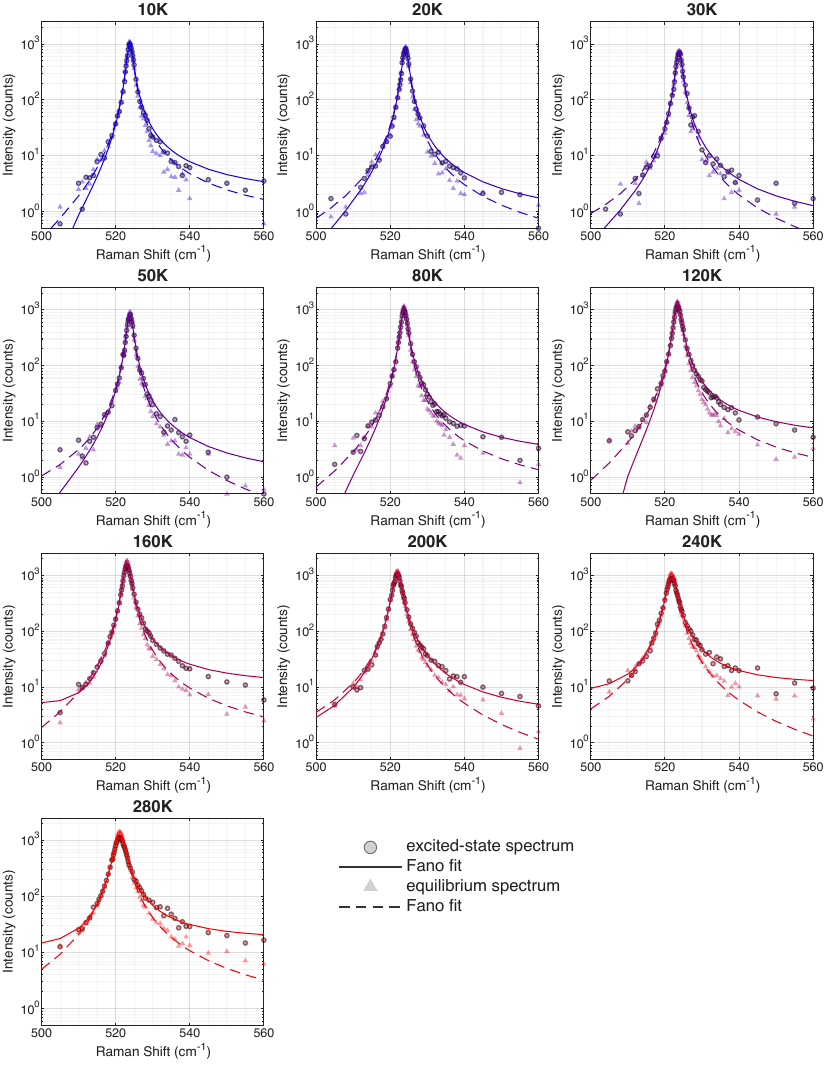}    \caption{\normalsize \justifying Temperature-dependent Fano lineshape fit results for the excited-state (circles) and equilibrium (crosses) spectra show a good fit.}
    \label{fig_SI:Fano_fit_Tdep_on_off}
\end{figure}

Figure~\ref{fig_SI:Fano_fit_Tdep_on_off} shows the excited-state and equilibrium Raman spectra alongside their fit to the Fano lineshape. 
The fit captures the data well; a slight peak asymmetry is visible when comparing the intensities of both peak tails.
The peak asymmetry is increasing with temperature, as expected from the increasing fraction of free charge carriers above 150~K due to thermal energy larger than the exciton binding energy~\cite{green2013improved}.

\begin{figure}[h!]
    \centering
    \includegraphics[width=0.9\linewidth]{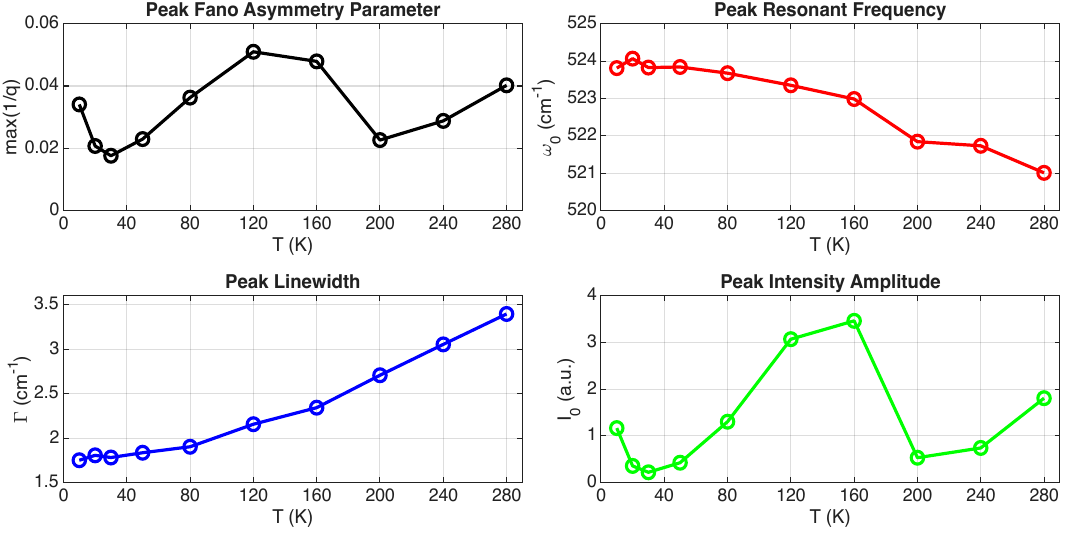}    \caption{\normalsize\justifying Temperature-dependent Fano fit parameters extracted at the excited-state spectrum.
    The q$^{-1}$ and I$_0$ temperature trends reveal physical anomalies.}
    \label{fig_SI:Fano_Tdep_excited_fit_params}
\end{figure}

The Fano fit parameters of the excited-state spectrum are extracted for each temperature, and plotted in figure~\ref{fig_SI:Fano_Tdep_excited_fit_params}.
The temperature trends of the resonance frequency $\omega_0$ and the linewidth $\Gamma$ are physically reasonable - the phonon resonance red-shifts and linewidth broadens with increasing temperature.
On the contrary, the temperature trends of the Fano asymmetry parameter $q^{-1}$ and the intensity prefactor $I_0$ are not.
The Fano asymmetry factor shows a non-monotonic trend, whereas the raw data, as well as previous reports on its temperature trend in equilibrium Raman scattering, showed a monotonic, increasing trend~\cite{Magidson2002Fano-typeSi}.
This is a confirmation that in transient photodoping, the Fano model is not physical when comparing different lattice temperatures, as it does not incorporate temperature effects~\cite{Balkanski1975TheorySemiconductors, Cerdeira1973EffectModes}.

\newpage

\subsubsection{Coupled-modes model analysis}
The coupled-modes (CM) model, based on the single-particle Green's function $G(\Omega)$~\cite{kwok1968green}, is employed to describe the interaction between vibrational modes in materials.
It is beneficial for systems where vibrational states are coupled to each other.
The model captures the coupling's impact on spectral features, including frequency shifts, linewidth changes, and the emergence of lineshape asymmetries.

\begin{figure}[h!]
    \centering
    \includegraphics[width=0.95\linewidth]{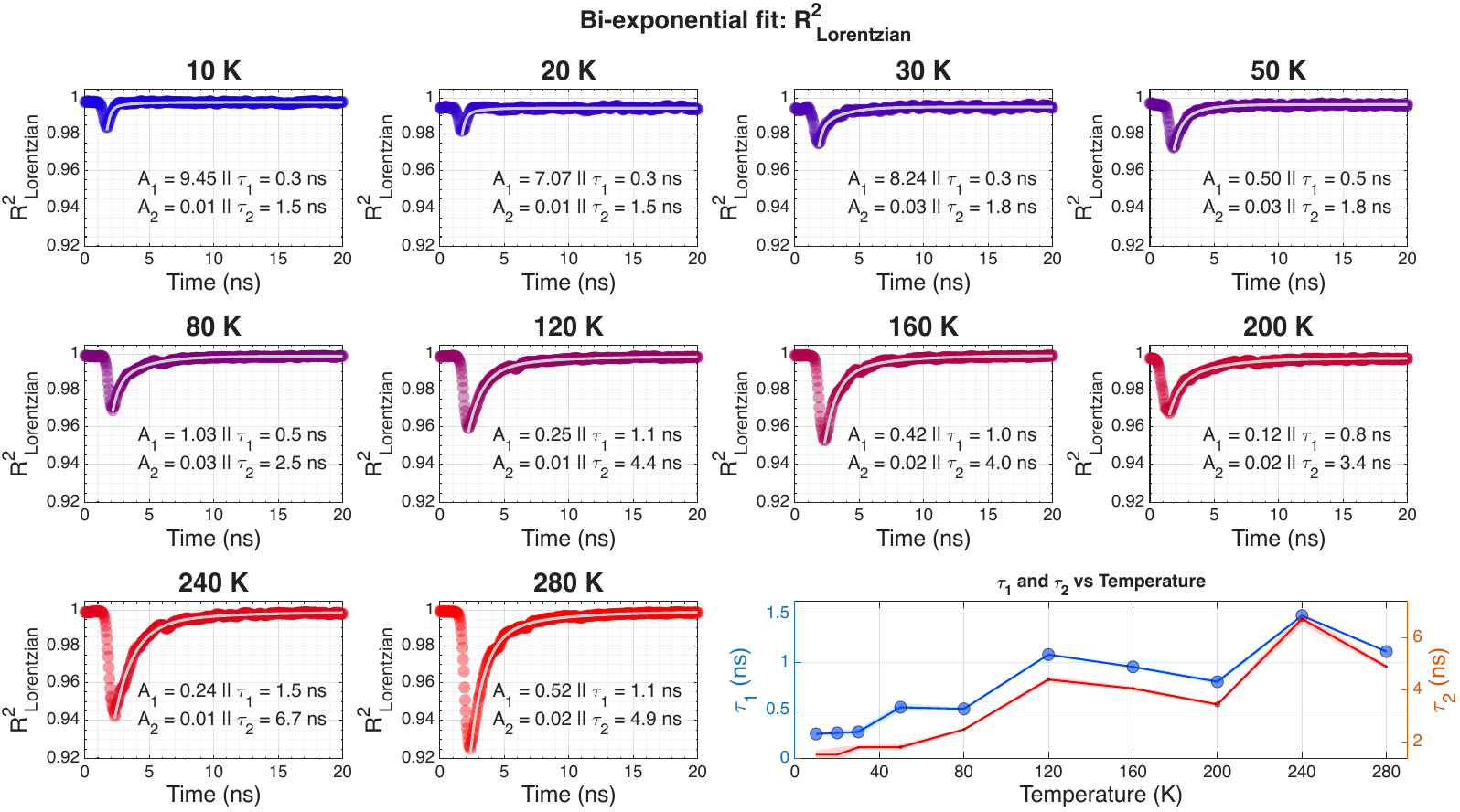}
    \caption{\normalsize \justifying Temperature- and time-dependent $R^2$ parameter of the Lorentzian lineshape fit tracks the temporal deviation of the spectrum from the equilibrium, Lorentzian lineshape. This quantity follows the population of excited charge carriers.
    All decaying curves were fitted to a bi-exponential decaying function.
    The bottom-right panel summarizes the time constants between 10--280~K for both $\tau_1$ and $\tau_2$.}
    \label{fig_SI:TCM_fit_R2_Tdep_transients}
\end{figure}

\begin{figure}[h!]
    \centering
    \includegraphics[width=0.95\linewidth]{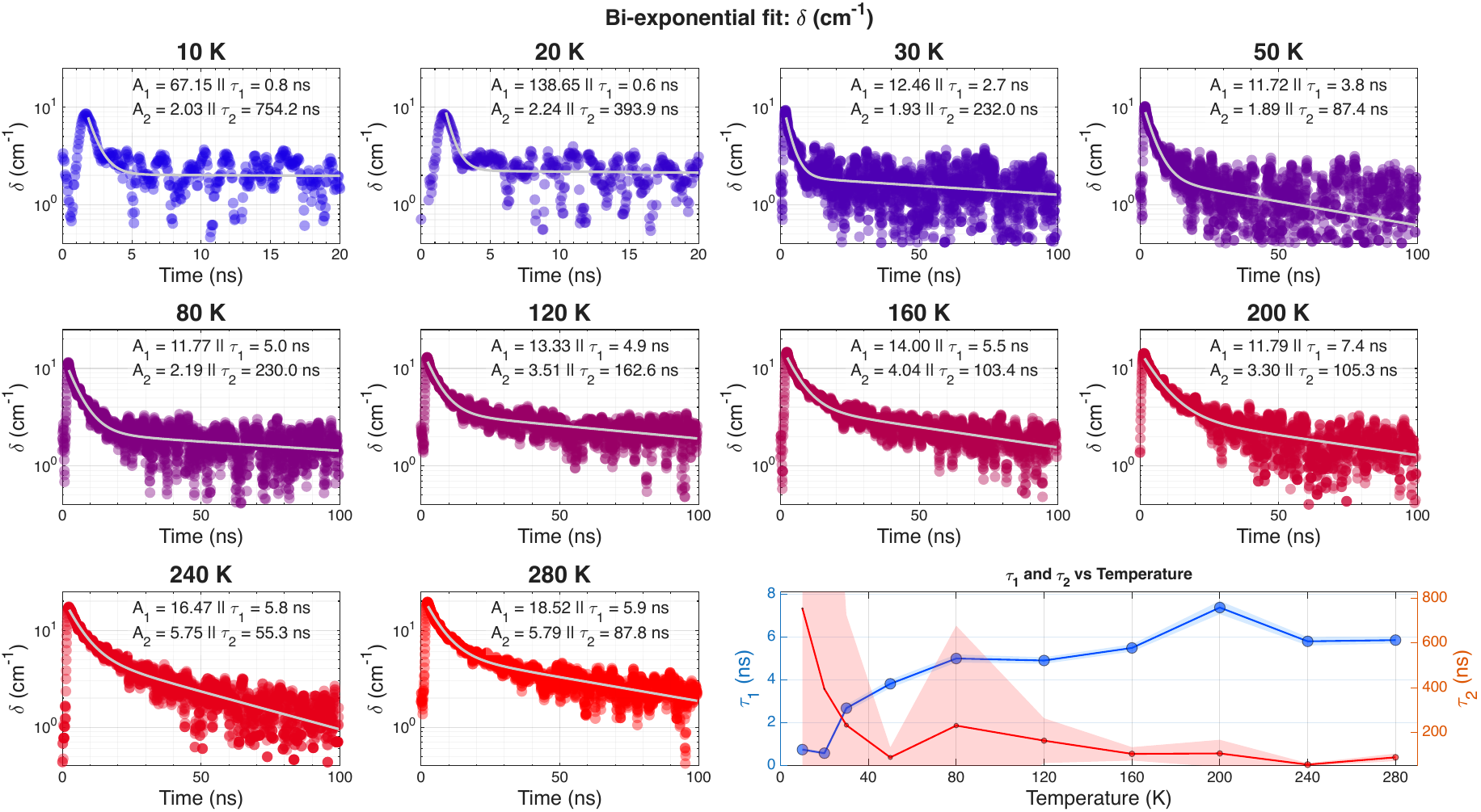}
    \caption{\normalsize \justifying Temperature- and time-dependent evolution of the coupled-modes fit parameter $\delta(t)$, the real part of the coupling term. 
    All decaying curves were fitted to a bi-exponential decaying function.
    The bottom-right panel summarizes the time constants between 10--280~K for both $\tau_1$ and $\tau_2$.}
    \label{fig_SI:TCM_fit_delta_Tdep_transients}
\end{figure}

\begin{figure}[h!]
    \centering
    \includegraphics[width=0.95\linewidth]{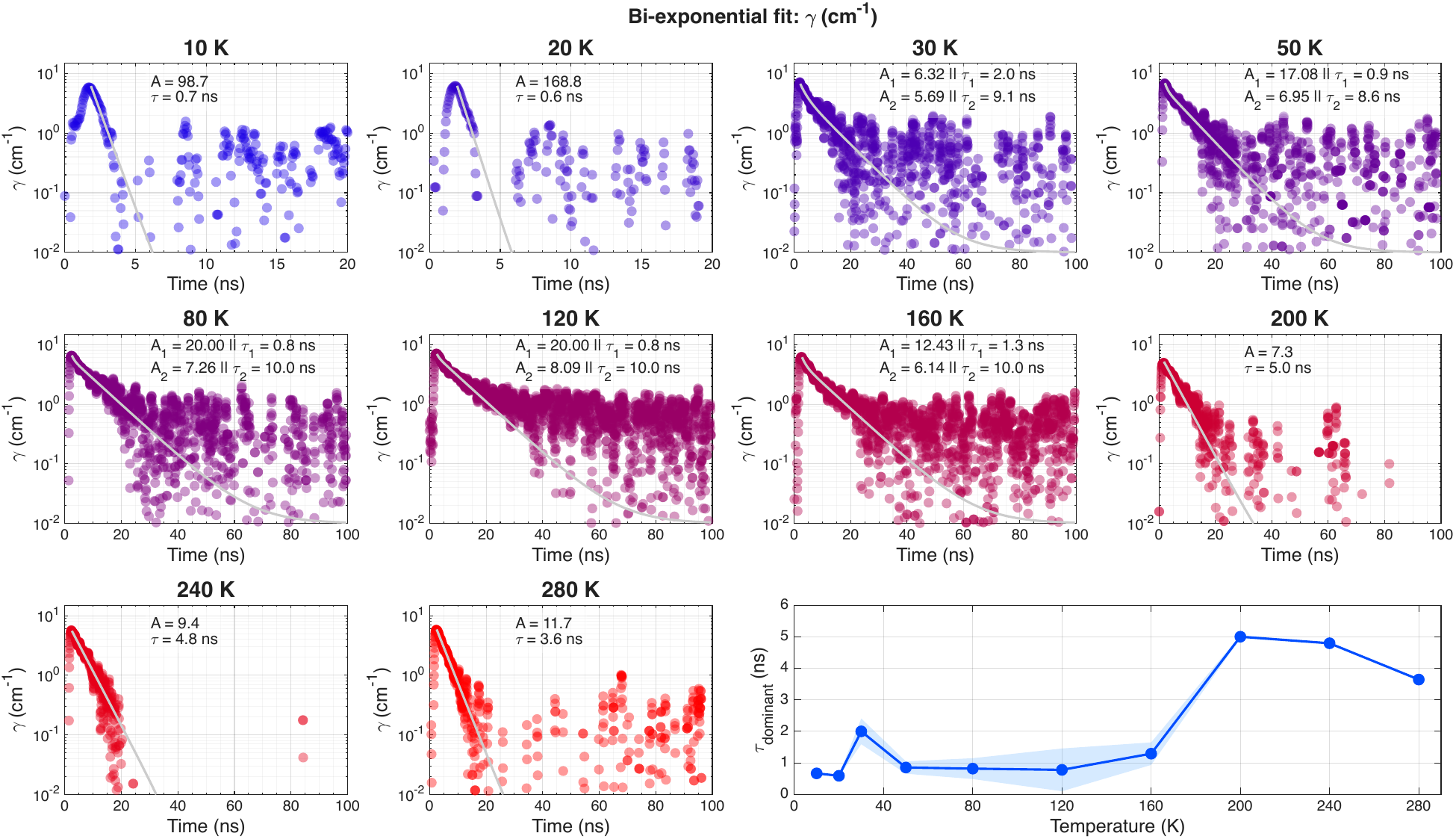}
    \caption{\normalsize \justifying Temperature- and time-dependent evolution of the coupled-modes fit parameter $\gamma(t)$, the imaginary part of the coupling term. 
    The decaying curves were fitted to either a single- or bi-exponential decaying function, according to the best fit, and the single or dominant exponent value in the bi-exp. curves were extracted to summarize the time constants between 10--280~K.}
    \label{fig_SI:TCM_fit_gamma_Tdep_transients}
\end{figure}

We vary the temperature of the silicon wafer and follow the variation of the CM fit parameters.
Figures~\ref{fig_SI:TCM_fit_R2_Tdep_transients} --  \ref{fig_SI:TCM_fit_gamma_Tdep_transients} show the temperature-dependent transients of each CM fit parameter, with a bi-exponential decay fit for its signal decay time. We note that only for the imaginary part of the coupling, $\gamma(t)$, both low and high temperatures fitted better to a single-exponential decaying function rather than a bi-exponential; therefore, the dominant exponent was chosen at the intermediate temperatures to plot the temperature trend.
As the imaginary part of the coupling term is related to the dissipation of the energy, it may have to do with the change in the dissipative channels at higher temperatures.
Notably, all fit parameters are decaying faster upon cooling, where $\tau_1$ values range between 0.3-1.5 ns ($R^2$), 0.6-7.4 ns ($\delta$), and 0.6-5.0 ns ($\gamma$).
The faster decay of the photoexcited effect is likely related to a change in the dominant quasiparticle dynamics associated with the formation of excitonic droplets in equivalent temperatures~\cite{forchel1982systematics, Revuelta2023, pelant2012luminescence}.
    
\clearpage

\end{document}